\documentclass[12pt]{article}
\usepackage{amssymb}
\usepackage{amsmath,bbm}
\usepackage{epsfig}
\usepackage{enumerate}
\def\ba{\begin{equation}}
\def\ea{\end{equation}}
\def\Bi{\begin{itemize}}
\def\Ei{\end{itemize}}
\def\bs{\bar s}
\def\ii{\item}
\def\bb{\begin{eqnarray}}
\def\ee{\end{eqnarray}}
\def\la{\label}
\def\nn{\nonumber \\}

\newcommand{\be}{\beta}
\newcommand{\pa}{\partial}
\newcommand{\f}{\frac}
\newcommand{\lam}{\lambda}
\newcommand{\g}{\gamma}

\newcommand{\de}{\delta}
\newcommand{\avg}[1]{\left\langle{#1}\right\rangle}

\def\as{\alpha_s}

\def\ka{\kappa}
\def\eps{\epsilon}
\def\d{\partial}
\def\de{\delta}
\def\al{\alpha}
\def\g{\gamma}
\def\f{\frac}

\def \bw{\bar w}
\def\taub{\bar \tau_R}
\usepackage{amsmath}
\usepackage{graphicx}

\begin{document} 
\title{On the maximal noise for stochastic and QCD traveling waves}
\author{Robi Peschanski\thanks{
e-mail: {\tt robi.peschanski@cea.fr}}\\
\small Institut de Physique Th\'eorique\\
\small Unit\'{e} de Recherche associ{\'e}e au CNRS\\
\small CEA-Saclay, F-91191 Gif/Yvette Cedex, France.}
\date{}
\maketitle

\begin{abstract} 
Using the relation of a set of nonlinear Langevin equations  with reaction-diffusion processes, we note the existence of a maximal strength of the noise for the stochastic traveling wave solutions of these equations. Its determination is obtained using  the field-theoretical analysis of  branching-annihilation random walks near the directed percolation transition. We study its consequence for  the stochastic Fisher-Kolmogorov-Petrovsky-Piscounov equation. For the related Langevin equation modeling the Quantum Chromodynamic  nonlinear evolution of the gluon density with rapidity, the physical maximal-noise limit may appear  before the directed percolation transition, due to a shift in the traveling-wave speed. In this regime, an exact solution is known from a coalescence process. Universality and other open problems and applications are discussed in the outlook.
\end{abstract}
 
\section{Introduction}
The aim of the present note is to discuss some aspects of traveling wave 
solutions of a class of nonlinear equations with noise. These nonlinear Langevin equations  are all related to the continuum limit of reaction-diffusion processes in statistical physics.

A typical exemple of this class of equations is the stochastic Fisher-Kolmogorov-Petrovsky-Piscounov (sFKPP) equation. This equation  
reads
\begin{equation}
\f{\pa w(x,t)}{\pa t}=D\  \f{\pa^2 w}{\pa x^2}+ \lam\ (w-w^2)+ \eps\ 
\sqrt{w(1-w)}\ \nu(x,t)\ ,
\label{eq:u}
\end{equation}
where the white noise verifies $\avg{\nu}=0$ and
\ba
\avg{\nu(x,t)\nu(x',t')} =  \de(x\!-\!x')\ \de(t\!-\!t')\ .
\label{eq:noisenu}
\end{equation}
The deterministic part of \eqref{eq:u} is  the well-known FKPP equation 
\cite{FKPP}. $D$ is the diffusion coefficient, $\lam$ characterizes the 
$activation$ strength since it is responsible for an exponential grow of the FKPP solution in the dilute regime (when $w^2 \ll w$). At some point the nonlinear damping term becomes effective when entering the dense regime. The fluctuation contribution which sensibly modifies the solutions as we shall discuss in length, is characterized by the value of $\eps,$  the 
noise strength. 

The same type of equation appears in various fields of statistical 
physics and, recently, in the domain of  Quantum Chromodynamics (QCD), the  interaction theory of quarks and gluons, where it happens to model \cite{fluct0} the evolution of the gluon momenta in the wave-function of a hadron or a nucleus when the energy increases.

The main property of the FKPP equation  \cite{wave} is to  admit ``universal'' traveling 
wave 
solutions. Indeed, one finds in a quite general way traveling-wave solutions of 
the form $w(x\!-\!vt)$ where the speed $v$ is   independent of the initial 
conditions (provided they are sharp enough in $x$). The reason is that a {\it critical} speed is selected after some evolution in time. Qualitatively, this 
property may be considered as  a  {\it critical} non-equilibrium mechanism since it corresponds to a solution flowing from an unstable $(w\!=\!0)$ to a stable $(w\!=\!1)$ fixed point. It
results from the ``strain'' between the exponential increase raised up by the 
initial conditions in the linear regime and the damping due to the 
nonlinear terms. Subasymptotic terms of the speed and the structure of the wave 
front in the so-called ``leading edge'' region \cite{wave1} can also be predicted. There are 
many results and applications of this traveling wave property in the literature 
\cite{wave1}.

In physical problems, one has  often to take into account the existence of 
fluctuations. One simple exemple is the application of an evolution equation 
for 
a discrete set of ``walkers'' on a  lattice generating statistical fluctuations at the edge, 
whose first approximation in the continuum limit is  a cut-off, $e.g.$ on the function $w$ 
of Eq.\eqref{eq:u} for the sFKPP equation. Indeed, it has been shown that, even with only a tiny 
cut-off \cite{BD}, the effect on the 
asymptotic solutions is sizeable. The effect of fluctuations in the dilute part 
of the 
system (near $w\!=\!0$) has been confirmed by simulations \cite{greg}. It is 
amplified for two reasons. If of negative sign, a random 
fluctuation can drive the system to $w\!=\!0$ where it stays indefinitely, and 
thus 
acting as an effective cut-off. If positive, the fluctuation is amplified by 
the 
exponential increase in the dilute region and extends its effect far into the 
dense region. Both effects have been analyzed and clearly identified in the 
weak-noise regime \cite{fluct}. In fact, the solution may be interpreted as a 
stochastic combination of traveling waves with some dispersion $\Delta$ and the 
average solution changes from a function of {\small ${x\!-\!vt}$} to a function of $\ \f{x-vt}{\Delta \sqrt 
t}.$ In particle physics it corresponds to a transition from ``geometrical 
scaling'' 
\cite{geom} to ``diffusive scaling'' \cite{diff} properties of the gluon momentum distribution.

In the strong-noise regime, the fluctuations are expected to have  a 
decisive effect in low-dimensional problems.  Indeed in the case of the sFKPP equation,  an exact solution can be found using a duality relation with reversible reaction processes at the coalescence limit \cite{limit,mueller}. It has also been applied to the QCD case  \cite{marquet}. In both cases, the exact solution  features noisy traveling waves and correlations fully dominated by the strongest fluctuations.

However, one may question  whether one can go to arbitrary strength of the 
noise. Indeed, as we shall see through the general relation of the Langevin equations with reaction-diffusion processes, one has to take into account the existence of a phase transition which belongs to the universality class of the directed percolation (dp). At some point, the system no more propagates. 

Hence we are led for our purpose to take into account the fluctuations in an appropriate framework. The study of fluctuations  is where the field-theoretical formalism is certainly useful, since they are properly taken into account  in the very definition of the action. Our aim is thus to investigate what new information the field-theoretical 
formalism of reaction-diffusion processes can bring to the traveling-wave 
problem for the class of Langevin  equations to which  sFKPP and the QCD evolution equations belong. The field-theoretical formalism for 
reaction-diffusion processes \cite{doi,CT,Tauber} has 
been already used in the context of the weak-noise regime of  traveling waves  \cite{PL}, where it confirmed the results of the cut-off analysis 
of Ref.\cite{BD}. Our aim in the present note is to use the  field-theoretical formalism in the full range of noise strength, and more precisely at the strong-noise limit, where renormalization effects allow for a proper account of the large fluctuation effects.

In fact the main tool used  of our note is to take benefit of a thorough 
field-theoretical analysis \cite{CT} of specific reaction diffusion process, the 
Branching Annihilation Random Walks (BARW),  which we will relate to the strong 
noise problem of the class of  Langevin equation we want to investigate. Indeed, the continuum limit of BARW processes leads to a Langevin equation of the same class. Hence the renormalization analysis of the BARW process leads to the existence of a maximal noise strength and to its evaluation. We will argue that this result extends to the class of langevin equations we study. However, the specific case of QCD will need a supplementary analysis, due to a different term in the corresponding Langevin equation.
 
The plan of our note is the following: In section 2, we briefly recall some 
aspects of the Langevin equations applied to  reaction-diffusion and  QCD  
processes and  in section 3, we infer the  existence of the maximum noise strength from the study of the BARW process and give an evaluation. In  section 4, we apply our result  
to the  two exemples of the sFKPP equation and  to the 
 QCD nonlinear evolution. We close this note in section 5 by a summary  of our results, with a discussion  on  the universality range of our result (for more general reaction-diffusion)  and on possible  applications to QCD. 
\section{Langevin equations and reaction-diffusion processes}

 \vspace{.1cm}
{\bf Field theoretical formalism of the Langevin equations}\\ \\
It is known  that  Langevin equations of sFKPP type can be analyzed 
in the form of a bosonic quantum field theory \cite{doi}. This formalism is 
particularly convenient to treat  fluctuations superimposed to mean-field 
equations of FKPP type, since they are intrinsically taken into account by 
specific terms in the field-theoretical action. For instance, the field theory 
corresponding to Eq.\eqref{eq:u} is defined in terms of bosonic fields $w$ and their auxiliary counterparts $\bw$ by the action
\ba
S[w,\bw]=\int \!d^dx\ 
dt\left\{\bw\left[\pa_t\!+\!D\nabla^2\right]w\!-\!\lam\ \bw 
w(1\!-\!w)\!-\!
\f{\eps^2}{ 2}\ \bw^2 
w(1\!-\!w)\right\}\ .
\la{action}
\ea
In fact 
these auxiliary fields appear as external source fields for the deterministic 
(for linear terms in $\bw$) and the noise terms (for quadratic terms in $\bw$). 
Let us illustrate  the field-theoretical formulation \cite{doi} with the example of \eqref{action}: performing the path 
integral over  the terms linear in the auxiliary field $\bw$ gives rise exactly to the mean-field terms which, in the case of \eqref{eq:u} is nothing else than the FKPP 
equation. Performing the gaussian path integral of  the quadratic part in $\bw^2$ allows one to  include the
fluctuation contribution  and generates  correspondingly the 
noise term in the sFKPP version \eqref{eq:u} of the Langevin formalism.

This field-theoretical framework casts a useful   bridge between the methods of quantum field theory and the solution of reaction 
diffusion processes and with the treatment of equations of the sFKPP type (Ref.\cite{CT} will be of particular interest for us, while for a comprehensive review, see \cite{Tauber}). Indeed, in such  reaction-diffusion processes it is 
possible to discuss  the probability of evolution from an ``inactive'' state to 
an ``active'' one, under the effect of  branching (or splitting), 
diffusion, annihilation and merging (or coagulation) of random walkers.  In the continuum limit, it contains in full generality\footnote{It is even more general since some  reaction-diffusion processes cannot be described by a consistent Langevin equation ( $e.g.$ generating an imaginary noise term).} the 
field-theoretical description of the 
Langevin  equations of the type  \eqref{eq:u}. We put 
this connection on a precise footing in the following sections.

One major interest of the field-theoretical treatment of the reaction-diffusion 
system is the  treatment of  fluctuations which play an important role in low-dimensional 
systems, such in our $d\!=\!1$ case. It 
appears in particular through the renormalization 
 of the field theory. In fact  due to fluctuations, a phase transition happens which is  in the 
universality class of directed percolation \cite{CT,Tauber}.
 \\

\vspace{.1cm}
{\bf Langevin Equation for QCD evolution processes}\\ 

\begin{figure}[t]
\includegraphics[width=13cm]{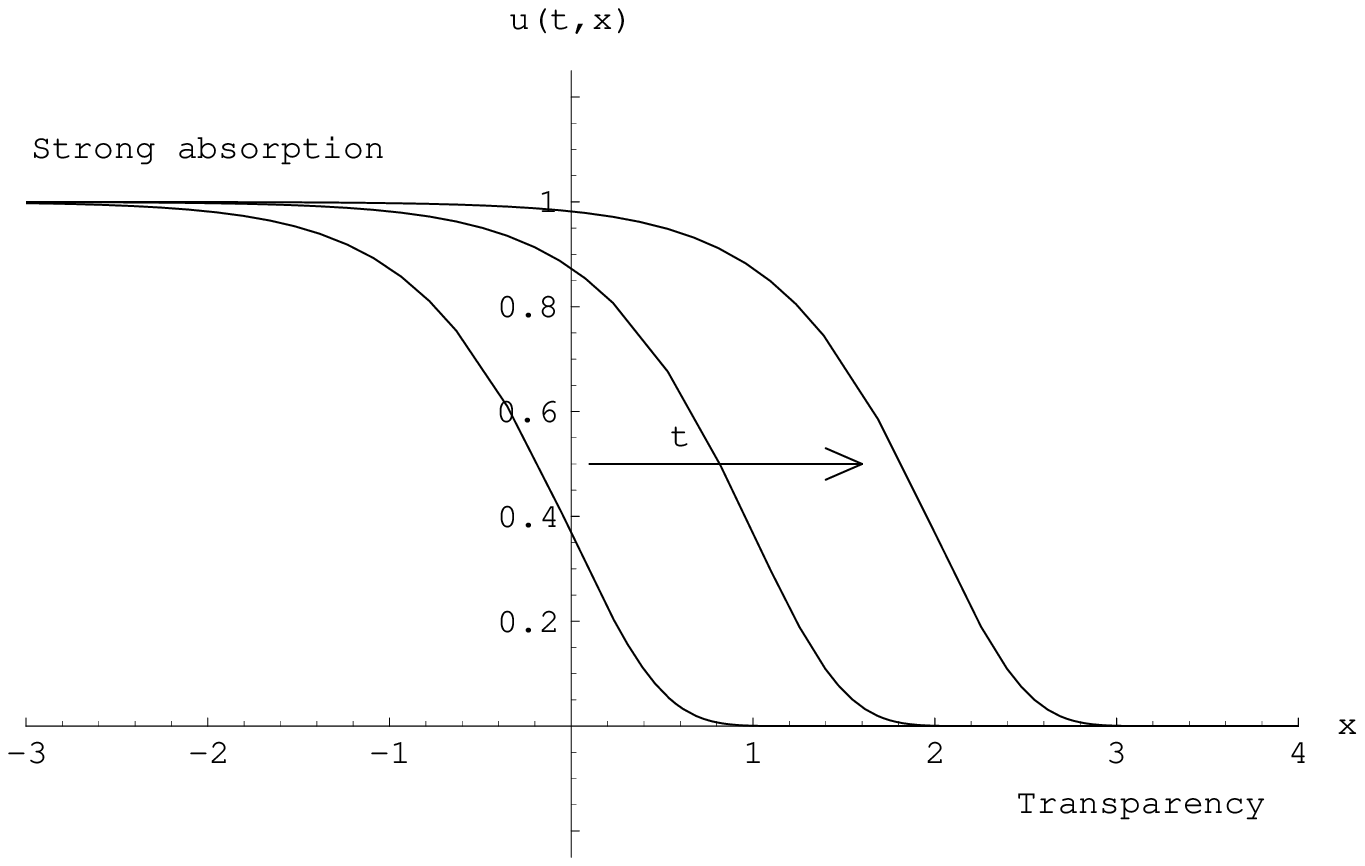}
\caption{{\it QCD traveling waves, from Ref.\cite{munier}}. 
\newline The traveling waves are mean field solutions of QCD evolution equations,  translationally invariant in time and joining the unstable fixed point(transparency or  {\it dilute} medium of gluons), to the stable fixed point (strong absorption or {\it dense} gluon medium).}
\label{1}
\end{figure}

In the  QCD problem the region near the 
``inactive'' state corresponds to the boundary of the dilute  region for the gluon density
(or ``transparency'' where the S-matrix  $S=1$) while 
the  ``active'' state corresponds to the boundary of the  dense region  (or ``strong absorption'' where the S-matrix  $S=0$),
  see Fig.\ref{1} and  
Ref.\cite{review} for a review.

The QCD evolution process which we refer to is thus the rapidity dependence of the 
tranverse-momentum 
gluon distribution in a nucleon or nucleus target \cite{review} which evolves from the transparency to the strong absorption regime when the rapidity ($i.e.$ the logarithm of the energy) increases. This equation,  at least within an approximate framework can be given a field-theoretical interpretation in terms of non-equilibrium processes.

Indeed, a coarse-graining approximation \cite{iancu} leads to a simplified description based 
on the stochastic 
equation
\ba
\partial_Y T(L,Y)  = \chi(-\partial_L)\ T(L,Y) -  T^2(L) + \sqrt{\ka \as^2  
T(L,Y)}\ \nu(L,Y)\ .
\label{eq:langevin}
\end{equation}
where $L=\log k^2,$  $k$ is the gluon transverse momentum, $Y$  the 
rapidity in 
units 
of the strong coupling 
constant $\as$ and $x$  the fraction of total energy brought by the 
gluon. We will consider $\chi$ as the leading order in $\as$   Balitsky-Fadin-Kuraev-Lipatov (BFKL) kernel \cite{bfkl}
\ba
\chi(\gamma)=2\psi(1)-\psi(\gamma)-\psi(1-\gamma)\ .
\label{eq:lkernel}
\end{equation}
Modifications by higher order corrections will not be introduced  at the present stage of our derivation  but they certainly  occur and deserve further study. 
The noise $\nu(L,Y)$ satisfies $\avg{\nu}=0$ and 
\ba
\avg{\nu(L,Y)\nu(L',Y')} = \f 2\pi \ \de(Y\!-\!Y')\ \de(L\!-\!L')\ .
\label{eq:noise}
\end{equation}
The  deterministic  part of  \eqref{eq:langevin} is the Balitsky-Kovchegov (BK)
equation   \cite{Balitsky}. Its solutions have been found to converge to 
traveling waves  after some rapidity evolution \cite{munier}. This fact is particularly clear if one uses the diffusive approximation as we shall see now. It leads  to a deterministic part of the equation identical to  FKPP with a modified noise term.

Indeed,
expanding the BFKL kernel  to second order around some $\gamma_c,$
\begin{equation}
\chi(\gamma) = \chi_c + 
\chi'_c(\gamma\!-\!\gamma_c)+\frac{1}{2}\chi''_c(\gamma\!-\!\gamma_c)^2
             = A_0 + A_1\gamma + A_2\gamma^2\ ,
\label{eq:coefsai}
\end{equation} 
The  value of $\g_c$ is obtained through the equation  $\g_c\ \chi'(\g_c)=\chi(\g_c)$ which determines the critical speed of the traveling waves. In fact, it can be shown \cite{munier} that $\g_c$ plays the role of an anomalous dimension characterizing   the wave front formed by the ``blocking'' mechanism described in introduction.

Introducing the \textit{space}   $(x)$ and \textit{time}  $(t)$  variables by the redefinition  
 \cite{munier} 
\begin{equation}
t = \as Y,\quad 
x = L\!-\!A_1 \as Y,\quad u(x,t)=\f{T}{A_0}
\ ,
\label{eq:space-time}
\end{equation} 
Eq.\eqref{eq:langevin} gets  mapped, within the diffusive approximation \eqref{eq:coefsai}, onto the Langevin equation:
\begin{equation}
\f{\pa u(x,t)}{\pa t}=A_2\  \f{\pa^2 u}{\pa x^2}+ A_0\ (u(x,t)-u^2(x,t))+ 
\f{2\as}{\pi}\sqrt{\kappa {A_0}\ u(x,t)}\  \nu(x,t)\ .
\label{eq:w}
\end{equation}

It is thus easy to realize that the first term 
$A_0$ corresponds to  branching, $A_2$ to the diffusion constant. 
The space derivative  term with coefficient $A_1$ in \eqref{eq:coefsai}, is thus   translated into  a time-dependent $shift$  on the space variable  by the 
redefinition \eqref{eq:space-time}  of the kinematic variables. This shift  plays, as we shall see, 
an important role in the application of reaction-diffusion processes to QCD and to the maximal noise problem in particular.

Hence, in the coarse-graining approximation of QCD evolution process, the deterministic 
part 
of \eqref{eq:w} can be interpreted as  gluon momentum diffusion, branching  and merging while 
the noise $\nu$ is a 
simplified description of the gluon-number fluctuation effects, which in the diagrammatic QCD framework is associated to the 
so-called ``Pomeron loops''. The strength of these fluctuations 
is proportional to $\as$ while the coarse-grained fudge factor $\ka$ may be large.

Writing
\ba
D \equiv A_2,\quad 
\lam \equiv A_0,\quad \eps \equiv \f{2\as}{\pi}\sqrt{\kappa {A_0}}\ ,
\la{couplings}
\ea
Eq.\eqref{eq:w} is similar to the 
sFKPP
equation \eqref{eq:u}, except for the quadratic term in the noise which is here absent. 

In  the Langevin formulation, the quadratic term  $-w^2$ in the noise of \eqref{eq:u} cancels the fluctuations near the stable fixed point of the mean-field equation at $w\!=\!1$, while they are not constrained near $u\!=\!1$ in \eqref{eq:w}. 
In the following discussion however, we shall consider  Eq.\eqref{eq:w} and \eqref{eq:u} together with the other  ones differing only by the quadratic term in the noise as belonging to the same class of equations. The restrictive condition will be that no unphysical fluctuations be generated in the range spanned by the mean-field solution, $i.e.$ the overall noise factor should stay positive in that range, such as for Eqs.(\ref{eq:u},\ref{eq:w}).

Once this condition satisfied, the fluctuations are in general expected to be small in the ``dense system'' and without strong effect while they have an important role at small $u$ or $w$ which is the region of the dilute system. We will focus on this region in our study, and thus conjecture that our result is valid for the whole abovementionned class of Langevin equations with different  quadratic term in the noise contribution. We will come back to this point in the discussion of our results.

In the following we will use the diffusive approximation of the QCD equation. 
However, one should keep in mind that a complete treatment of the fluctuations  beyond the coarse-grained and fixed coupling approximations is still an open problem, since higher-order QCD contributions may change the features of the solution\footnote{The effect of introducing  the running of the QCD coupling constant in the QCD evolution equations with noise has been recently studied either through a numerical modelization \cite{Dumitru:2007ew} or on the BK equation with noise \cite{beuf}. It seems that the fluctuation effects are transferred to higher rapidities than with fixed coupling.}.

\section{Branching-Annihilating Random Walks}
As already mentionned the existence of a critical activation 
parameter has been found in BARW \cite{CT} together with an estimate 
of its value within a 
\newline $2\!-\!d$ expansion at first order. An  
improved value could be found with higher order corrections or within the exact renormalization group approach 
\cite{leonie}. We postpone this evaluation to further work.

For sake of generality, let us start by  considering a reaction-diffusion process in dimension $d$, including the  
transition processes  
$A\to A+A,$ with rate $\lam,$ $A+A \to (0,A)$ with rates 
$(\mu,\mu')$ respectively, and  $A_i\to A_{i\pm1}$ (where {\tiny $i\pm1$} is 
a short-hand notation for  nearest neighbours in dimension $d$) with diffusion rate $D$.
One  writes the following 
action \cite{Tauber} for the probability distribution $s$ of the species $A$
\ba
S[s,\bs]=\!\int d^dx\   
dt\left\{\bs\left(\pa_t\!-\!D\nabla^2\right)s\!-\!\lam_{\cal R} \bs s \!-\!\lam\bs^2\!s+\!(2\mu\!+\!\mu')\bs s^2 \! +\! (\mu 
\!+\!\mu')\bs^2 
s^2\right\}
\la{action1}
\ea
where $\lam_{\cal R}$ takes into account the extra process $A\to 0$ which is anyway imposed by  renormalization of the $A\to A+A.$ 
 A general study of the reaction-diffusion action \eqref{action1} in terms of $4-d$ expansion  
\cite{Tauber} is possible but the critical dimension being $d_c=4,$ (with the noticable exception of the BARW process) the study we perform in dimension $d=1$ is not trivial. We will discuss further this point in the last section.

Let us then consider more specifically  the BARW transition processes, which has the interesting special feature \cite{CT} to possess a critical dimension  $d_c=2.$ The BARW processes are described by the 
action \eqref{action1} restricted to $\mu'=0,$  $i.e.$ no coagulation process $A+A\to A$. Using the same formalism 
\cite{Tauber,PL}, and considering for the present being the bare action ($\lam_{\cal R}=\lam$), 
the theory  corresponds to the  Langevin equation
\ba
 \f{\partial s}{\partial t}= D \nabla^2 s +\lam  s - 2\mu\ s^2 
+\sqrt{2\lam s-2\mu s^2}\ \eta(x,t)\ ,
\la{eq:B}
\ea
This Langevin equation is well-defined $i.e.$ the noise factor is real 
for $s\le \lam/\mu,$ while the ``active state'', which is the maximum value of 
$s$ in the mean-field approximation, is twice  as small as the value $\lam/2\mu$ where the noise is again zero. The noise is non zero at the stable fixed point $s=1$ of the mean-field equation  but only very exceptional fluctuations could cause positivity problems under the square root. As already mentionned, we expect this feature not to be  relevant for our study. The Eq.\eqref{eq:B} belongs to the same class of Langevin equations as  \eqref{eq:u} and \eqref{eq:w} that we have considered, and thus the results obtained with the BARW process will be expected to be valid for the other cases. Through the redefinition 
\ba
s= \f{\lam}{2\mu}\ w\ ;\quad \eps=\sqrt{4\mu}
\la{noiseBARW}
\ea
we indeed recover the familiar form of these equations, namely the same as \eqref{eq:u},  up to the quadratic term in the noise being now $\eps\ \sqrt{w-\f {w^2}2}\ \nu(x,t).$

In Ref.\cite{CT}, the fluctations are shown to renormalize the activation 
parameter $\lam$ through the one-loop effect $A\to A+A\to 0,$ implying a counter-term in the action \eqref{eq:B} corresponding to the new transition $A\to  0.$ As one may infer, the renormalization contributes with a negative sign, showing that 
the fluctuations tend to stabilize the classically unstable  fixed 
point. 

Let us   now show that the minimum  obtained for the activation term in Ref.\cite{CT} 
corresponds to a maximal noise and evaluate its value.

After field-theoretical renormalization \cite{CT} using the $2\!-\!d$  expansion at first order, one finds 
the renormalized 
value $\lam_R$ of the activation parameter
\ba
\lam_R=\lam\ \left\{\f {1-\f{\mu}{2\pi(2\!-\!d)}\ 
(\lam D)^{-\!1\!+d/{ 2}}}
{1+\f{\mu}{2\pi(2\!-\!d)}\ (\lam D)^{-\!1\!+d/{ 2}}}\right\}\ .
\la{renorm}
\ea
This induces, when $\lam_R=0$  and $ d \le 2$ a {\it minimum} value 
\ba
\lam_{min}= \f 1D \left(\f{\mu}{2\pi}\right)^{\f 2{2-d}}
\la{mind}
\ea
below which there is no 
transition from the inactive to the active state. In terms of the directed percolation phase transition which is met at this point, the system no more percolates.

From the point of view of  the Langevin formalism  of \eqref{eq:B} in  $d=1$ and its  solutions, it means that for 
\ba
\lam_{min}= \f 1D\  {\left(\f{\mu}{2\pi}\right)^{2}}
\la{min}
\ea
there is an extinction of the noisy traveling waves. 

Considering now  the dimensionless noise  $\tilde \eps\equiv 
{\eps}/{\left(D\lam\right)^{1/4}},$ as defined in \cite{mueller},  and the 
parametric relations \eqref{noiseBARW}, one finds: 
\ba
\tilde \eps \equiv \f{\eps}{\left(D\lam\right)^{1/4}} \ 
\le  \ \tilde\eps_{max} ={ \left(\f{{16\mu^2}}{D\lam_{min}}\right)^{1/4}} =\ \sqrt{8\pi} 
\ .
\la{basic6}
\ea

In fact, the key point of this maximal noise property is that when the noise is strong enough, the inactive phase $(s=0)$ from being originally unstable in the mean-field or with  noise  weaker than the maximal one   becomes {\it stable} when the noise strength reaches the value \eqref{basic6}. This prevents the instability which leads to the noisy traveling waves going towards the stable fixed point. This {\it stabilisation} by the noise is a phenomenon appearing  under various forms in statistical physics (see $e.g.$ \cite{kirone}).

Note that the value \eqref{basic6} has been obtained from the evaluation \cite{CT} of the 
minimum activation parameter at first order in the $2\!-\!d$ expansion of the 
renormalization procedure. A more precise estimate could be probably obtained 
from higher-order calculations or from the exact renormalization group analysis \cite{leonie}. The best hope would be an exact solution in $d=1.$

\begin{figure}[htb]
\hspace{-3cm}\includegraphics[width=17cm]{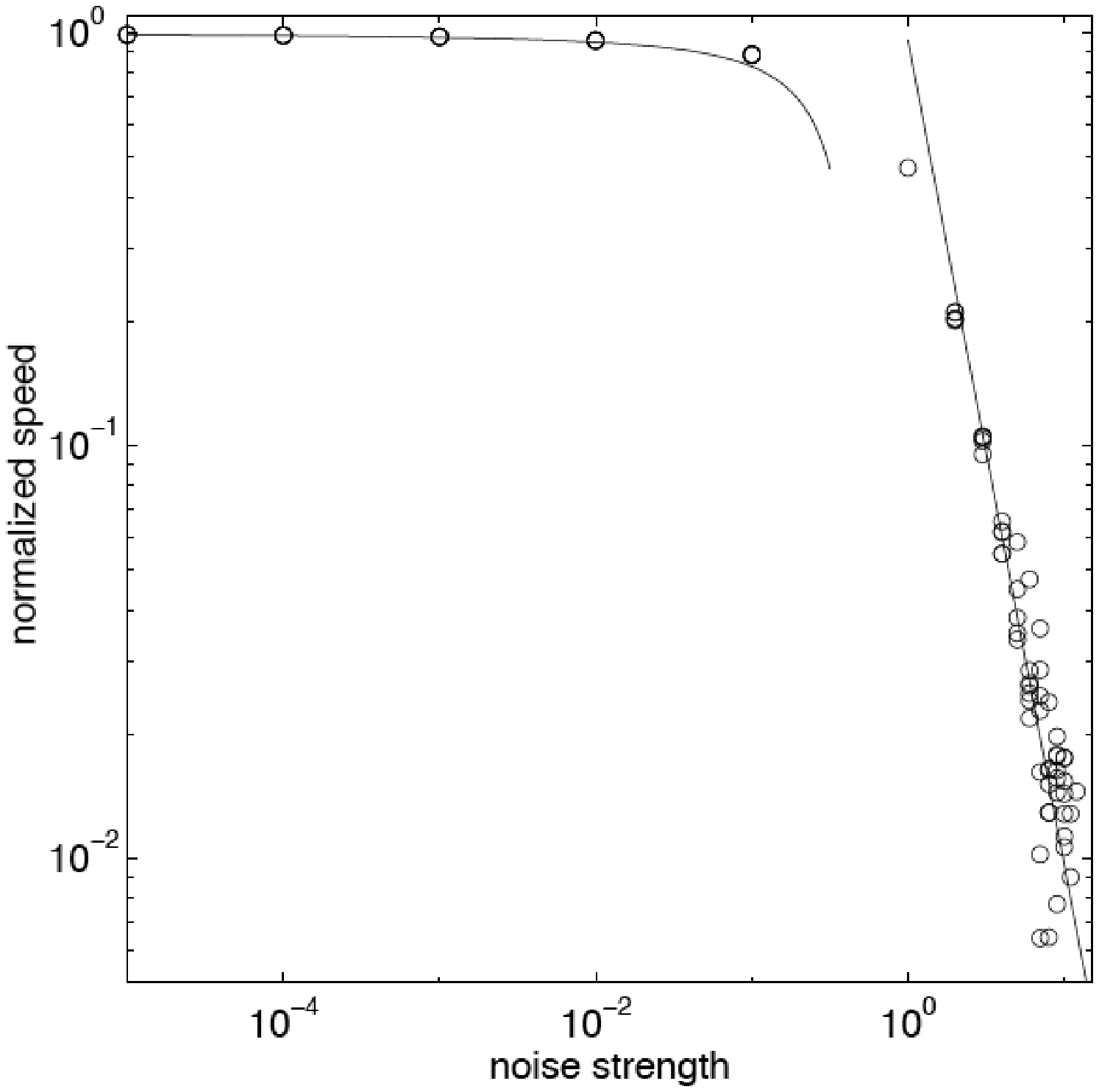}
\caption{{\it Wave speed as a function of the noise strength for the solutions of the 
sFKPP equation, from ref.\cite{mueller}}. 
\newline Vertical axis: traveling wave speed 
normalized to the critical one $2\sqrt{D\lam}$ of the deterministic FKPP 
equation. Horizontal axis: dimensionless noise strength  $\bar \eps 
=\eps/(D\lam)^{\f 14}.$ Line on the left: the weak noise prediction 
\cite{BD,PL}; Line on the right: the strong noise prediction \cite{limit}. One 
observes a maximal speed around a noise strength of order 10.}
\la{2}
\end{figure}

\section{Applications to  sFKPP and QCD} 
\subsection{Maximal noise for sFKPP traveling waves}

In Ref.\cite{mueller}, the sFKPP equation is studied by using duality 
properties 
with the $A\leftrightarrow A+A$ reaction-diffusion process with  creation or recombinaison, each with its own rate. At strong coupling, this dual
reaction-diffusion
process goes to a limiting $coalescence$ regime, where at most one walker occupies a lattice site and with a probability of giving birth to another one in a neighbouring site.  This limit allows one to find exact analytic solutions  \cite{limit}. The 
resulting average wave  speed, in terms of 
$v_c=2\sqrt{D\lam}$  the speed in the absence of noise \cite{BD,PL} and $\tilde\eps= \eps/(D\lam)^{\f 14}$  the dimensionless 
noise strength  \eqref{basic6} takes the form \cite{mueller}.
\ba
 v=\f{2D\lam}{\eps^2}= {2\sqrt{D\lam}}\times 
\f{\sqrt{D\lam}}{\eps^2}\ =\ \f{v_c}{\tilde \eps^2}\ .
\la{diff}\ea

The comparison with numerical simulations has been  done in Ref.\cite{mueller} 
and reproduced in  
Fig.2. It shows the following features: in the range of weak noise, the 
normalized
speed agrees quite well with the theoretical predictions 
\cite{BD}. With larger noise strength, one can observe that, when ${\cal O}(1)\le \tilde \eps \le {\cal O}(5\!-10),$
the numerically measured  speed  agrees reasonably well, within some 
statistical spread, with the theoretical prediction using the coalescence
approximation. 

However the end of the spectrum, not discussed in Ref.\cite{mueller}, seems to occur, compatible with a maximal noise limit. Indeed, the average speed of the numerical solution appears to stop near, indeed slightly superior,  to the $2\!-\!d$ expansion value $\sqrt{8\pi}\sim 5,$ see Fig.2.  We thus note that the 
approximate value of ${\tilde\eps}_{max}$ is within the range of our 
estimate from the BARW process. It is an interesting challenge bpth from the numerical and theoretical point of view, to check whether more refined evaluations could confirm this finding. We hope to improve the this value by going beyond the  1-loop result or  $via$ the exact renormalization-group approach \cite{leonie}.

This  region of noise strength below but near the transition where  fluctuations stabilize the  unstable mean-field fixed point corresponds to the 
critical region where the diffusion-reaction system meets the critical line of a phase transition of the directed percolation class \cite{CT,leonie}. Hence,
beyond the determination of the end-point of the noise strength, it would  be interesting to discuss the other  predictions coming from 
the renormalization  properties at the phase transition \cite{Tauber}, for instance  the relevance of the critical exponents to the description of the sFKPP solution and the  matching with the near-by $coalescence$ limit approximation of the related reaction-diffusion process. This study is certainly deserved for the future.

\subsection{Maximal noise for QCD traveling waves} 

Coming back to the problem of QCD traveling waves with strong fluctuations, we have already noticed (see \eqref{eq:space-time}) that in
the diffusive approximation of the QCD 
equation there is a change of  space variable 
\ba
x = L\!-\!A_1 \as Y \la{qcd1}\ ,
\ea
where $A_1$ is the (negative) coefficient of the kernel expansion 
\eqref{eq:coefsai}. 

Hence, in the absence of noise, Eq.\eqref{qcd1} amounts to a shift for the wave speed
 \ba
{v_{QCD}^{no\ noise}} = {v_c}\ \left(1- \f{|A_1|}{2\sqrt{A_0 A_2}}\right)\la{qcd}\ ,
\ea 
with respect to the critical value $v_c=2\sqrt{A_0 
A_2}.$ 

In the noisy regime dual to  $coalescence$ one finds \cite{marquet} 
 \ba
\f{v_{QCD}^{strong\ noise}}{v_c} = \f 1{(\tilde \eps)^{{ 2}}} - \f{|A_1|}{2\sqrt {A_2 A_0}}\ 
\le \f 1{(\tilde \eps_{max})^{{ 2}}} - \f{|A_1|}{2\sqrt {A_2 A_0}},
\la{sqcd}\ea
where we have given the limit corresponding to the maximal sFKPP noise. 
Following \eqref{sqcd} and the expression \eqref{basic6} of the maximal 
dimensionless noise strength, the normalized speed of the average QCD wave 
depends critically on the value of the ratio  $\rho=\f{|A_1|}{2\sqrt{A_0 A_2}}$ 
compared to the maximal noise strength  $\tilde \eps_{max}.$
\begin{itemize}
\item If $\tilde \eps^{-{ 2}}_{max} \le \rho \le 1:$ The QCD speed goes to zero for a value of the 
noise strength which has not reached the value corresponding to \eqref{basic6}; The system  does not reach the phase transition for positive speed (required by the physical interpretation of a rapidity evolving gluon distribution.
\item  $\rho \le \tilde \eps^{-{ 2}}_{max} :$ The system reaches the critical point where 
appears 
 the  directed percolation phase transition where the wave traveling stops. The behaviour of the solution near the endpoint should be given by the renormalization group analysis.
\end{itemize}

In fact, if one sticks to the QCD leading-log order BFKL kernel \eqref{eq:lkernel} and the value 
$\tilde\eps_{max} \approx 5\!-\!10$ extracted from the simulations \cite {mueller}, 
one sees that we are in the case of the first item:  the directed percolation transition point 
cannot be reached. Indeed, in 
 Ref.\cite{marquet}, two options were discussed for the diffusive 
approximation of the kernel for the strong-noise region, depending on which 
value of $\g_c$ the kernel has to be developped up to the second derivative. If 
one considers the   same critical value $\g_c$ as in the no noise case, one 
finds with $(A_0,A_1,A_2)=(9.55,-25.6,24.3)$ and thus $\rho\sim .9$ . If one considers a 
small value of 
$\g_c \ll 1$ , one finds $(A_0,A_1,A_2)\sim (3\g_c^{-1},-3\g_c^{-{2}},1\g_c^{-3})$ 
and 
thus $\rho=\sqrt{3}/2,$ independently of $\g_c$. In both cases we find $\rho 
\approx .8\sim.9 < \tilde\eps_{max} \approx 5\!-\!10$,
and thus, referring to Fig.\ref{2},  the system of QCD traveling waves ends (reaching $v_{QCD}=0$) within the
coalescence regime, without reaching the sFKPP maximal noise\footnote{The result could change with corrections from higher QCD pertubative orders but would need  a profound modification of the kernel. One could also consider yet unknown strong $\as$ coupling effects.}.

In order to give some appropriate estimate, using  relation \eqref{couplings}, and writing $\eps < \eps^{QCD}_{max},$ one finds the relation
\ba \f{2\as}{\pi}\sqrt{\kappa} <  
\left(\f{2 A_2}{A_0}\right)^{\f 12} \ .
\la{noiseQCD}
\ea 
Hence, for $\f{A_0}{A_2}\sim {\cal O}(1)$ the  maximal noise 
 could be reached even in the perturbative regime 
($\f{2\as}{\pi}$ small) if the fudge factor $\kappa$ is large enough. On the other hand, it is  interesting to consider the effect of increasing  the coupling constant $\as$. We see that in our framework it leads to an increase of the noise and consequently a decrease of the wave speed. Such an effect is otherwise expected in QCD when considering the transition towards the non-perturbatice regime. Both considerations  give 
an incentive to  developping further a coalescence model starting from  the original QCD formukation $e.g.$ directly in terms of branching annihilation and merging of gluons or QCD dipoles.

\section{Summary of results and discussion}

In the present note, which used some known results on the field-theoretical 
treatment of stochastic Langevin equations of sFKPP type and of related 
reaction-diffusion systems, we have found the existence of a limiting maximum 
strength of fluctuations, which can be quantified by a maximum value of the 
noise parameter in the Langevin equations. Below this value, there exists 
noisy traveling wave solutions which describe the transition of the system 
from an ``inactive'' phase to an ``active'' phase. Beyond this point, the 
fluctuations are strong enough to prevent the system to undergo such a 
transition and it stays in the inactive phase, which is otherwise unstable. Hence, we have yet another striking example of random fluctuations contributing to stabilize a deterministic unstable point. 

Pursuing the analogies with the associated 
reaction-diffusion systems, the critical value of the noise corresponds to a 
directed percolation critical point, where the systems ceased to percolate. In this framework the noisy traveling waves could be interpreted as an average over the percolation path histories. 

We have discussed two applications of the results. In the sFKPP case, we 
interpreted the existing numerical simulations showing the compatibility with 
the existence of a maximal noise strength in the range predicted by the 
field-theoretical calculations. For traveling waves in the diffusive 
approximation of the QCD evolution equations, the traveling waves go to zero 
before the transition point of the sFKPP analoguous equation due to a shift of 
the speed (at least for the leading order kernel in $\as$). The maximal noise-strength occurs in the region related by duality 
with a  $coalescence$, which is 
accessible to exact noisy traveling wave solutions, as discussed in \cite{marquet}. For large enough QCD fudge factor, it may be 
reached even in the perturbative QCD regime. Otherwise, it may indicate the slowing down of rapidity evolution expected from the nonperturbative QCD regime.

An important aspect of the discussion is the extent to which our result on the maximal noise is  ``universal'', that is valid for a whole class of processes.

  In our study we have met different
 Langevin equations, namely Eq.\eqref{eq:u}  corresponding to sFKPP, Eq.\eqref{eq:w} corresponding to the coarse-graining of the QCD evolution equation in the diffusive approximation and finally Eq.\eqref{eq:B}  corresponding to the BARW process and for which we could extract the value of the maximal noise. Since these Langevin equations are all well defined and differ only by a quadratic term in the noise which is not relevant for  the region near the unstable mean-field fixed point, we have conjectured that the result on the maximal noise should  be the same, and thus given  by \eqref{basic6} in the first-order approximation of the $2\!-\!d$ expansion for the BARW process. 

Since each of these Langevin equations corresponds to the continuum limit of a specific reaction-diffusion process with different parameters for merging, splitting and annihilation, it is interesting to address directly the universality property in terms of a renormalization problem of the field theories for these three cases.  By extension, one may investigate the validity of the obtained maximal noise strength for all reaction-diffusion processes pertaining to the universality class of the directed percolation processes \cite{Tauber}, for the class of processes
for which the  formulation in terms of a Langevin equation differs only by a quadratic term in the noise (unless leading to imaginary noise term), which is irrelevant near the directed percolation transition point.

This task is not easy in the framework of the $d_c\!-\!d$ expansion since the critical dimension is  $d_c\!=\!4$ in the directed percolation class, which is the universality class of the phase transition of these reaction-diffusion processes. It is a long way towards $d\!=\!1.$ However, a qualitative general argument can be given which relies on the topological property of random walks in dimension $d<2.$ Indeed, in that condition, any two random walks will have probability 1 to meet again. Hence, if  the combination of reaction-diffusion constants contribute to a large enough dimensionless noise strength, one expects a stabilization of the otherwise unstable fixed-point. Starting with the general expression \eqref{action1}, and following a similar discussion as for the  BARW case, 
one is led to conjecture
$\tilde\eps_{max}= \sqrt{8\pi}\ +$ (higher order contributions).
It is thus deserving  to  check this  result
 in $d\!=\!1$\footnote{Note added in proof; The special case $d=0$ has been reconsidered very recently. In Ref.\cite{recent}, the authors examine the diffusion-reaction systems with various couplings in the zero transverse dimension approximation. They find that only the reversible process $A\to A+A,$ $A+A\to A$ leads to an increase of the solution with time (or the amplitude with rapidity in QCD language). They conclude that only the sFKPP equation (in its reduction to zero transverse dimension) to which this equation is related gives a satisfactory toy model of high-energy QCD. We note that in our case $d\!=\!1,$ the diffusion term (absent in zero transverse dimension) is crucial for the properties of the process, especially in the dilute region. For instance the reduced noise $\tilde \eps$ becomes infinite at zero diffusion constant. The properties reported in \cite{recent} seeem to be related with the noise in the region of the other mean-field fixed point (and thus  depending critically on the quadratic term in the noise expression, contrary to the $d\!=\!1$ case), which is leading to negative evolution in time, except for the reduction of the sFKPP equation in $d=0.$}.

As an outlook one could propose two distinct directions for future work. In the 
sFKPP case, it would be useful to develop the connection between the regime just 
below the maximal noise with the results obtained by the renormalization group method near the  directed percolation critical point. In 
particular the knowledge of the critical exponents should give interesting 
predictions for the behaviour of the averaged traveling wave and its eventual 
deformation at the critical point. From the QCD point of view, the existence of 
a maximal noise-strength at least 10 time smaller than the sFKPP one, leads to 
interesting predictions from the coalescence process $via$  duality properties \cite{limit,mueller}. It would be 
interesting to develop on the already known studies \cite{marquet} by looking to 
 dual properties similar to the coalescence limit more directly  from the fromulation of QCD evolution processes in terms of merging, splitting and annihilation of gluons or QCD dipoles.
 
\section*{Acknowledgements}
We thank Peter Smereka for allowing us to reproduce the Fig.\ref{2} containing the numerical results of Ref.\cite{mueller}. We are indepted to Guillaume 
Beuf, Cyrille Marquet, Kirone Mallick and  Gregory Soyez for  careful reading of the manuscript, many useful suggestions and for constructive criticisms. Discussions with L\'eonie Canet and Bertrand Delamotte are acknowledged. We thank Jos\'e Guilherme Milhano for communicating the Ref.\cite{recent} prior to publication.

\end{document}